\documentclass{aa}
\usepackage{psfig}
\begin{document}

   \thesaurus{08     
              (02.04.1;  
               02.13.2;  
               08.09.3;  
               08.13.1;  
               08.14.1;  
               08.16.6)} 
   \title{Diquark condensates and the magnetic field of pulsars
	\thanks{Research supported in part by the Volkswagen Stiftung under
grant no.\ I/71 226}}


   \author{D. Blaschke \inst{1}
        \and D. M. Sedrakian\inst{2}
	\and K. M. Shahabasyan\inst{2} 
          }

   \offprints{D. Blaschke}

   \institute{Fachbereich Physik, Universit\"at Rostock,
	Universit\"atsplatz 1, D--18051 Rostock, Germany\\
	 email: blaschke@darss.mpg.uni-rostock.de 
        \and Department of Physics, Yerevan State University, Alex
	Manoogian Str. 1, 375025 Yerevan, Armenia
             }

   \date{Received April 28, 1999; accepted September 19, 1999}

   \maketitle

   \begin{abstract}

We study the consequences of superconducting quark cores in neutron 
stars for the magnetic fields of pulsars.
We find that within recent nonperturbative approaches to the effective quark 
interaction the diquark condensate forms a superconductor of second kind 
whereas previously quark matter was considered as a first kind superconductor. 
In both cases the magnetic field which is generated in the surrounding hadronic
shell of superfluid neutrons and superconducting 
protons can penetrate into the quark matter core 
since it is concentrated in proton vortex clusters where the field strength 
exceeds the critical value. 
Therefore the magnetic field will not be expelled from the
superconducting quark core with the consequence that there is no decay of the 
magnetic fields of pulsars.
Thus we conclude that the occurence of a superconducting quark matter core in 
pulsars does not contradict the observational data which indicate that 
magnetic fields of pulsars have life times larger than $10^7$ years.   

   \end{abstract}

%


Recently, the possible formation of diquark condensates in QCD at finite 
density has been reinvestigated in a series of papers following Refs. 
(\cite{arw98}; \cite{r+98}).
It has been shown that in chiral quark models with a nonperturbative 
4-point interaction motivated from instantons (\cite{cd98}) or nonperturbative 
gluon propagators (\cite{br98}, \cite{brs99}) the anomalous quark pair 
amplitudes in the color antitriplet channel can be very large: of the order 
$\approx 100~ {\rm MeV}$.
Therefore, in two-flavor QCD, one expects this diquark condensate to dominate 
the physics at densities beyond 
the deconfinement/chiral restoration transition and below the critical 
temperature ($\approx 50~ {\rm MeV} $) for the occurence of this 
``color superconductivity'' (2SC) phase. 
In a three-flavor theory it has been found (\cite{arw99}, \cite{sw99}) that 
there can exist a color-flavor locked (CFL) phase 
for not too large strange quark masses (\cite{abr99})
where color superconductivity is 
complete in the sense that diquark condensation produces a gap for quarks of 
all three colors and flavors, which is of the same order of magnitude as that
in the two-flavor case. 

The high-density phases of QCD at low temperatures are most relevant for 
the explanation of phenomena in rotating compact stars - pulsars. 
Conversely, the physical properties of these objects (as far as they 
are measured) could constrain our hypotheses about the state of matter at the 
extremes of densities. 
In contrast to the situation for the cooling behaviour of
compact stars (\cite{bkv99}) where the CFL phase is dramatically different 
from the 2SC phase, we don't expect qualitative changes of the magnetic 
field structure between these two phases. Consequently, we will restrict 
ourselves here to the discussion of the simpler two-flavor theory first.

According to Bailin and Love (1984) the magnetic field of pulsars should be 
expelled from the superconducting interior of the star due to the Meissner 
effect and decay subsequently within $\approx 10^4$ years.
If their arguments would hold in general, the observation of lifetimes for 
the magnetic field as large as $10^7$ years (\cite{pines}; \cite{bpp69}) would 
exclude the occurence of an extended superconducting quark matter phase in 
pulsars.  
For their estimate, they used a perturbative gluon propagator which yielded a
very small pairing gap and they made the assumption of a homogeneous magnetic 
field. 
Since both assumptions seem not to be valid in general, we perform a 
reinvestigation of the question whether presently available knowledge about 
the lifetime of magnetic fields of pulsars might contradict the occurence
of a color superconducting phase of QCD at high densities.


The free energy density in the superconducting quark matter phase with 
$ud$ diquark pairing ($J^P=0^+$ and color antitriplet index $p$) is given by
(\cite{bl84}) 
\begin{eqnarray}
f&=&f_n + \alpha d_{p}^*d_{p} + \frac{1}{2}{\beta}(d_{p}^*d_{p})^2 \nonumber\\ 
&& + \gamma(\nabla d_{p}^* +iq\vec{A}d_{p}^*)(\nabla d_{p} -iq\vec{A}d_{p})
+\frac{B^2}{8 \pi}~,
\end{eqnarray}   
where $\vec{B}={\rm rot}\vec{A}$ is the magnetic induction, $q$ the charge of 
the 
$ud$ pair and the coefficients of the free energy are given by the following
expressions
\begin{eqnarray}
\alpha&=&\frac{d n}{d E} t~~,\nonumber\\
{\beta}&=& \frac{d n}{d E} \frac{7 \zeta (3)}{8 (\pi k_{\rm B} T_{\rm c})^2}~~,
\nonumber\\
\gamma &=& \frac{d n}{d E} \frac{7 \zeta (3)}{48 (\pi k_{\rm B} T_{\rm c})^2}
\frac{p_{\rm F}^2}{\mu^2}=\frac{1}{6}\frac{p_{\rm F}^2}{\mu^2}{\beta}~~,
\end{eqnarray}
where $t=(T-T_{\rm c})/T_{\rm c}$. 
Here $T_{\rm c}$ is the critical temperature, 
$p_{\rm F}$ the quark Fermi momentum,
$\mu=\sqrt{p_{\rm F}^2+m^2}$ 
is the chemical potential (in zeroth order with respect 
to the coupling constant), $d n/d E=\mu p_{\rm F}/\pi^2$.

The Ginzburg-Landau equations for relativistic superconducting quarks are 
obtained in the usual way
\begin{eqnarray}
0&=&\alpha d_{p}+\beta(d_{p}^*d_{p})d_{p}+\gamma(-i\nabla-q\vec{A})^2 d_{p}~,
\nonumber\\
\vec{j}&=& 
iq\gamma(d_{p}\nabla d_{p}^*-d_{p}^*\nabla d_{p})-
2q^2\gamma d_{p}d_{p}^*\vec{A}~.
\label{gl}
\end{eqnarray}
In deriving the expression for the current $\vec{j}$ we have also used the
Maxwell equation
\begin{equation}
{\rm rot} \vec{B}= 4\pi \vec{j}~.
\end{equation}
The first of the Ginzburg-Landau equations (\ref{gl}) has a solution which 
corresponds to the Meissner effect ($\nabla d_{p}=0,~\vec{A}=0$ inside of the
superconductor):
\begin{equation}
\Delta^2=
|d_{p}|^2=-\frac{\alpha}{\beta}
= - \frac{8 t (\pi k_{\rm B} T_{\rm c})^2}{7 \zeta(3)}~.
\end{equation}
For the case of weak fields ($H<H_{c2}$) one obtains the London equation:
\begin{equation}
\vec{B} +\lambda_{\rm q}^2 {\rm rot~ rot} \vec{B} = 0~~,
\end{equation}
where $\lambda_{\rm q}$ 
is the penetration depth of the magnetic field into the 
superconducting quark condensate.
The region of the change of the order parameter $d_{p}$ can also be determined 
from (\ref{gl}) via
\begin{equation}
\xi_{\rm q}^2=-\frac{\gamma}{\alpha}
=-\frac{7 \zeta(3)}{48 t (\pi k_{\rm B} T_{\rm c})^2}
\left(\frac{p_{\rm F}}{\mu}\right)^2~.
\end{equation}
\begin{figure}[htb]
\psfig{figure=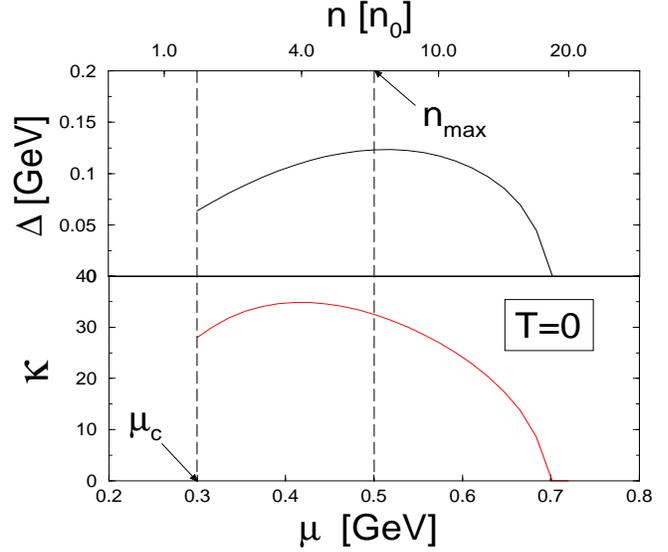,width=8.5cm,height=7.5cm,angle=0}
\caption{The dependence of diquark energy gap $\Delta$ and Ginzburg-Landau 
parameter $\kappa$  on chemical potential $\mu$  and  density n for a NJL-type 
quark interaction (\protect\cite{br99}); 
n$_0=0.16$ fm$^{-3}$ is the nuclear saturation density. 
The left dashed line denotes the critical chemical potential of the onset of 
quark superconductivity (the corresponding baryon number densities are given
on the upper scale), the right dashed line - the maximal values of chemical 
potential and density for stable stellar configurations.\label{kappa}}
\end{figure}
In the diquark condensate phase with a nonperturbative interaction 
 the energy gap is 
$\Delta
\approx 100$ MeV at $\mu \approx 400$ MeV \footnote{within a dynamical 
confining quark model the diquark gaps can be even larger than this estimate
(\cite{br98}).}, see Fig. \ref{kappa}. 
We obtain for the coherence length 
$\xi_{\rm q} = 0.8 \times 10^{-13}$ cm.
For the penetration depth of the magnetic field we have
\begin{equation}
\lambda_{\rm q}=\frac{1}{\sqrt{8\pi\gamma} q d_0}
=\sqrt{\frac{-3\pi\mu}{4 q^2 t p_{\rm F}^3}}
\approx 3.6\times 10^{-12} {\rm cm}~.
\end{equation}
The thermodynamical critical field $H_{\rm cm}$
that fully destroys the superconducting state in the case of a superconductor
of the first kind is given (\cite{bl84})
\begin{equation}
H_{\rm cm}^2=\frac{32 \pi \mu p_{\rm F} (k_{\rm B}T_{\rm c}t)^2}{7 \zeta(3)}~. 
\end{equation}
For the parameter values given above the critical field is
 $H_{\rm cm}\approx 8.7 \times 10^{17}$ G, i.e.  by two 
orders of magnitude larger than in Ref. (\cite{bl84}).

The Ginzburg-Landau parameter $\kappa$ which determines the behaviour of the
superconductor in an external magnetic field is given by (\cite{bl84})
\begin{equation}
\kappa=\frac{\lambda_{\rm q}}{\xi_{\rm q}}
=\sqrt{\frac{{\beta}}{8\pi \gamma^2 q^2}}
=132 \frac{\Delta}{\mu}\left(\frac{\mu}{p_{\rm F}}\right)^{5/2}~.
\end{equation}
For values of $\mu\sim p_{\rm F}\sim 400$ MeV/c and $\Delta=100$ MeV we obtain
$\kappa = 34$, see also Fig. 1. 
Therefore the superconducting quark condensate  appears as a 
superconductor of the second kind into which the external magnetic field can
penetrate by forming quantized vortex lines in the interval $H_{c1}<H<H_{c2}$.
The upper critical field $H_{c2}$ is determined by
\begin{equation}
H^{\rm q}_{c2}=-\frac{\alpha}{q \gamma}
=\frac{6 \Delta^2}{q}\left(\frac{\mu}{p_{\rm F}}\right)^2 
\approx 3 \times 10^{19} {\rm G}~.
\end{equation} 
The magnetic flux of the quark vortex lines $\Phi_{\rm q}$ amounts to
\begin{equation}
\Phi_{\rm q}=\frac{2 \pi \hbar c}{q}= \frac{2 \pi \hbar c}{e/3} 
= 6\frac{\pi \hbar c}{e} = 6 \Phi_0~~,
\end{equation}
where $\Phi_0=2\times 10^{-7}$ G cm$^2$ is the quantum of the proton magnetic 
flux.
The lower critical field for the occurence of quark vortex lines is then
\begin{eqnarray}
H_{c1}^{\rm q}=
\frac{\Phi_{\rm q}}{6\pi\lambda_{\rm q}^2}
\ln\frac{\lambda_{\rm q}}{\xi_{\rm q}}
= 1.8 \times 10^{16} {\rm G}~.
\end{eqnarray}
Here we have taken into account the spherical shape of the quark core in a 
neutron star (\cite{ssm84}).

Now we can describe the magnetic structure of a superconducting quark
condensate in a pulsar and its time evolution.
When during the cooling of the protoneutron star with a quark matter core 
the critical temperature for the transition to the 
superconducting state is reached in the presence of a magnetic field, then 
this field remains in the quark phase in the form of quantized vortex lines. 

At some point in the further rapid cooling of the star due to neutrino emission
the neutrons in the hadronic phase (``npe''-phase) of the star 
become superfluid.
Since the basic interaction between isolated protons resembles that of the 
neutrons, the protons in the hadronic phase become superfluid too. 
Since the density of protons in "npe"-phase is only few per cent of
the neutron density, the protons will pair in  $^1$S$_0$ pairing state
(\cite{ccy72}; \cite{ao85}; \cite{bcll92}).
The neutrons take part in the rotation, forming a lattice of quantized vortex
lines. Because of the strong interaction of the neutrons with the protons a 
part of the superconducting protons will be entrained by the neutrons 
(\cite{ss80}; \cite{als84}) and create in the region of the neutron 
vortex a magnetic field of strength $H(r)$ given by (\cite{ss80}; \cite{ssm83}) 
\begin{equation}
{H}(r)=\hat{\nu}_{\rm n}\frac{k \Phi_0}{2\pi\lambda_{\rm p}^2}\ln\frac{b}{r}~,
\end{equation}
where $b=\sqrt{\pi \hbar/\sqrt{3}m_{\rm n} \Omega}$ 
is the lattice spacing of the 
neutron vortex lattice, $k=(m_{\rm p}^*-m_{\rm p})/m_{\rm p}$ 
is the entrainment coefficient
with the effective mass $m_{\rm p}^*$ and the bare mass $m_{\rm p}$ 
of the protons;
$\hat{\nu}_{\rm n}$ is the unit vector in the direction of the vortex axis, 
$r$ is the distance from the center of the vortex and $\Omega$ is the angular
velocity of the rotation of the star.
This field, whose magnitude is determined by the rotation of the star, 
acts as an external field for the non-entrained protons and creates a cluster 
of proton vortices with the fluxes $\Phi_0$
in the region around the axis of the neutron vortex where 
$H(r)>H_{c1}^{\rm p}$.
The radius of this region, $\delta_{\rm n}$, equals (\cite{ssm84})
\begin{equation}
\delta_{\rm n}=b ({\xi_{\rm p}}/{\lambda_{\rm p}})^{\frac{1}{3|k|}}~.
\end{equation}
For the pulsar Vela PSR 0833-45 with $\Omega=70$ rad~s$^{-1}$ and 
$b=10^{-3}$ cm, we have $\delta_{\rm n}=10^{-5}$ cm. 

While the mean magnetic induction in the star due to proton vortex clusters 
is of the order $10^{12}$ G, the mean magnetic induction within the cluster 
reaches values of $4 \times 10^{14}$ G (\cite{ss91}; \cite{ss95}).

The magnetic field strength $H(r)$ which occurs in the ``npe''-phase is the 
strength of the external field relative to the superconducting quark 
condensate. 
It reaches the maximum value $H(0)$ close to the center of the neutron vortex,
i.e.
\begin{equation}
H(0)=\frac{k \Phi_0}{2 \pi \lambda_{\rm p}^2} \ln \frac{b}{\xi_{\rm n}}
\approx 4.7 \times 10^{16} {\rm G}~,
\end{equation}
for $\lambda_{\rm p}=30$ fm, $k=0.7$ 
and a coherence length $\xi_{\rm n}= 30$ fm of the neutron.

This external field generates quark vortex lines when the 
condition $H(r)\ge H_{c1}^{\rm q}$ is fulfilled.
The radius of this region is
\begin{equation}
\delta_{\rm q}=b({\xi_{\rm q}}/{\lambda_{\rm q}})^{\frac{2}{k}
(\lambda_{\rm p}/\lambda_{\rm q})^2}=4.3 \times 10^{-7} {~\rm cm}~~.
\end{equation}
This way, the entrainment current generates a strongly
inhomogeneous magnetic structure in the quark condensate:
the clusters of quark vortex lines with the fluxes $\Phi_{\rm q}$
and radii $\delta_{\rm q}$, the axes of 
which are the continuation into the quark phase of the axes of the neutron
vortex lines. Since $\delta_{\rm q}$ is by two orders of magnitude smaller 
than $\delta_{\rm n}$, the mean magnetic induction in the clusters of quark 
vortex lines increases to a value of the order of $ 10^{18}$ G.

When the condition for the applicability of the London approximation 
$H(0)\ll H_{c2}^{\rm q}$ is fulfilled, then one can apply the modified London 
equation
\begin{equation}
\vec{B}+ \lambda_{\rm q}^2 {\rm rot ~rot} \vec{B}
= \Phi_{\rm q} \hat{\nu}_{\rm q} \sum 
\delta(\vec{r}-\vec{r}_{\rm q})
\end{equation} 
for the description of the magnetic structure of the quark condensate. 
The density of clusters is equal to the density of neutron vortex lines
$n_V=2~\Omega/\kappa_{\rm n}$, where $\kappa_{\rm n}=\pi\hbar/m_{\rm n}$ 
is the quantum of neutron circulation.

\begin{figure}
\psfig{figure=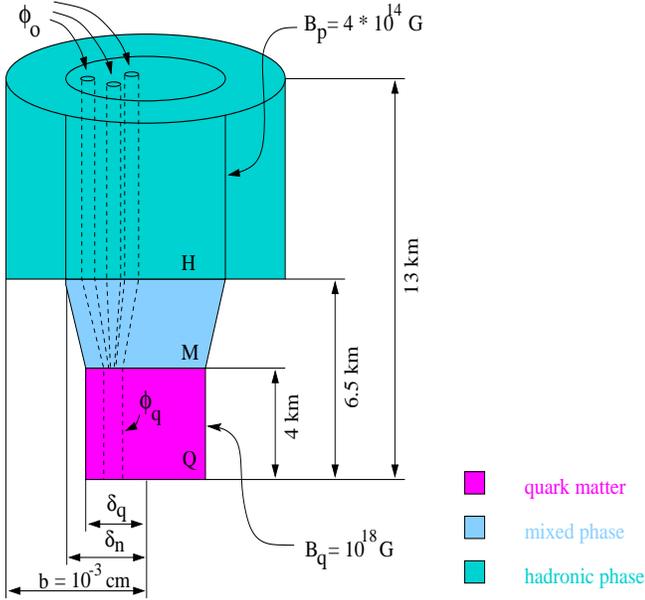,width=8.5cm,height=8cm,angle=0}
\caption{Magnetic field structure in the interior of a  hybrid star with 
$M=1.4 M_{\odot}$; $b$ is the radius of the neutron vortex
$\delta_{\rm n}= 10^{-5}$ cm is the radius of the proton vortex cluster, and 
$\delta_{\rm q} = 4.3 \times 10^{-7}$ cm is that of the quark vortex cluster. 
For details, see text.
\label{vortex}}
\end{figure}

We note that between the hadronic phase and quark core there is a mixed 
phase, in which hadrons coexist with a charged lattice of quark  droplets 
(\cite{ng92}).
Since the number densities of neutrons and protons in the mixed state are 
lower than in the hadronic phase, these particles remain superfluid.  
So neutron vortices and clusters of proton vortex lines continue through 
the mixed phase.
Therefore the magnetic field will pass through it and enter the quark core,
see Fig. \ref{vortex}.
In the case of small diquark gaps of the order of 1 MeV (\cite{bl84}), when the
diquark condensate is a superconductor of the first kind, the magnetic field 
generated in the ``npe''-phase penetrates into the quark matter core in the 
form of ordinary cylindrical regions (\cite{ssz97}). 
The radii of these regions will be of the order of $\delta_{\rm q}$ since the 
thermodynamical critical field $H_{\rm cm}$ is of the same order as the mean 
magnetic field in the quark cluster.
The clusters of quark vortex lines which appear due to the entrainment effect
in the ``npe''-phase will interact with those which are formed by the initial 
magnetic field (fossil field). 
This interaction obviously implies that quark vortex
lines will not be expelled from the quark core of the star within a time scale
of $\tau=10^4$ years as suggested in (\cite{bl84}).

We note that the evolution of the magnetic field is intimately related
to the rotational history of the star. In particular, the magnetic field of 
the quark core will decay because of the outward motion of neutron vortices 
when the star spins down. 
This behavior results from the fact that the magnetic clusters inside the 
quark core are the continuation of neutron vortices.
Therefore the characteristic decay time of the magnetic field for
the whole star (and also for quark core) is comparable to the pulsar's
slowing down time, which corresponds to the life time of the pulsar.

In conclusion, we find that the occurence of a superconducting quark matter 
core in pulsars does not contradict the observational data which indicate that
magnetic fields of pulsars have life times larger than $10^7$ years 
(\cite{pines}).
This holds true for small diquark gaps of the order of $1$ MeV (\cite{bl84}) as
well as for larger ones as obtained recently (\cite{arw98}; \cite{r+98}; 
\cite{cd98}; \cite{br98}) using effective 
models for the nonperturbative quark-quark interaction.

\begin{acknowledgements}
K.M.S. and D.M.S.  acknowledge the hospitality of the Department of 
Physics at the University of Rostock where this
research has been started. 
We thank H. Grigorian, K. Rajagopal, G. R\"opke, A.D. Sedrakian and 
D.N. Voskresensky for their discussions and comments. 
\end{acknowledgements}

\end{document}